\documentstyle[twoside,fleqn,epsfig,espcrc2]{article}

\newcommand{\AmS}{{\protect\the\textfont2
   A\kern-.1667em\lower.5ex\hbox{M}\kern-.125emS}}

\title{Jet Production at HERA}

\author{Sascha Caron                      
         (on behalf of the H1 and ZEUS collaborations)
        \\ 1. Phys. Institut, RWTH Aachen, Germany\\
        E-mail: scaron@mail.desy.de}                       

\begin{document}

\begin{abstract}
This article reviews recent jet physics results from HERA.
\end{abstract}

\maketitle

\section{Introduction}
Jet production allows precise tests of perturbative QCD.
At the electron-proton collider HERA
the production of jets is investigated in inelastic electron
(positron) proton scattering over a wide range of photon
virtualities $Q^2$, from photoproduction ($Q^2\approx 0 \, \mbox{GeV}^2$) to deep
inelastic scattering (DIS).
This article presents recent measurements of 
jet cross sections and jet shapes and their comparison 
to perturbative QCD (pQCD) calculations.
These calculations are performed up to next-to-leading 
order (NLO) for all measurements presented.
Jet cross sections at HERA are calculated by 
convoluting the short distance subprocess
cross sections with parton density functions 
(pdfs) of the proton.
At very small $Q^2$ (photoproduction)
the photon may fluctuate into partons before 
the hard scatter. This is parametrised by the
pdfs of the photon.

The standard choice since about 5 years at HERA is
 to define jets according to the theoretically preferred invariant $k_{\bot}$
algorithm as proposed by Ellis and Soper\cite{jet}.
ZEUS and H1 have reduced their energy scale uncertainty for
high transverse energy ($E_T$) jets 
to $1\%$, respectively $2\%$, which significantly improves the
data precision.
                                                                  
The data at high $Q^2$ and high $E_T$ allows quantitative tests of
pQCD and extractions of $\alpha_s$ and the
 proton pdfs.  Photoproduction data at high $E_T$ are, in addition, 
sensitive to the photon pdfs.
At moderate $Q^2$ and $E_T$ the limits of NLO pQCD are exploited.
These measurements provide additional and
complementary information to
$p\overline{p}$ and $e^+e^-$ data.

\section{Jets at moderate and high $Q^2$ DIS}
\begin{figure}[htb]
{\epsfig{file=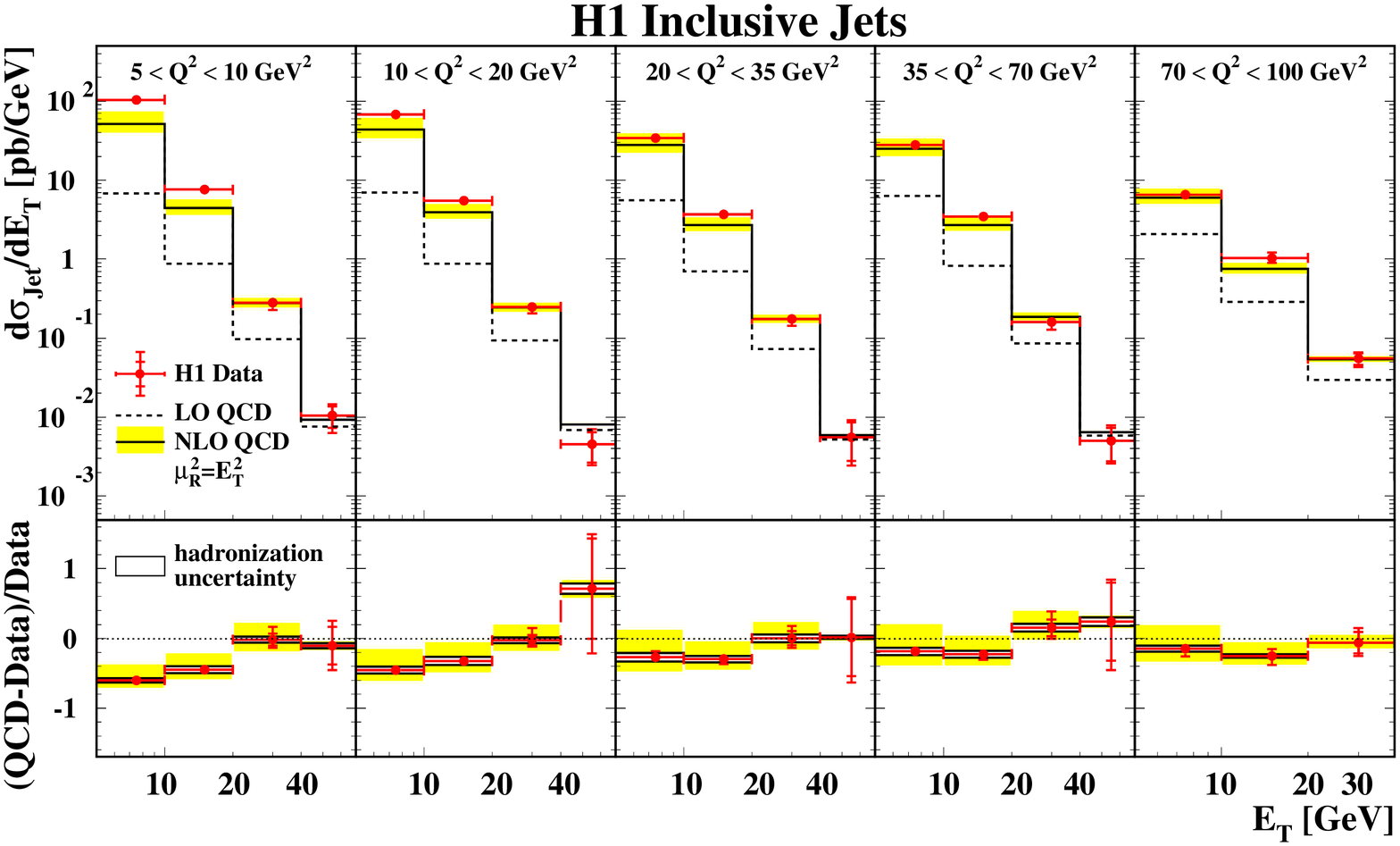,width=8cm}}  
\caption
{Differential
 $ep$ cross section $d\sigma/dE_T^{Jet}$ for inclusive jet production.
 The LO and NLO predictions are shown as dotted and dashed lines.
The lower figures show the relative difference between the predictions
and the data.
}
\label{fig1}
\end{figure}

Jet production in deep inelastic scattering events has been
investigated over a wide range in $Q^2$. A 
recent H1 measurement~\cite{Adloff:2002ew} of inclusive 
jet production is performed for
the medium $Q^2$ range between 5 and 100 $\mbox{GeV}^2$.
The data are compared to LO and NLO pQCD calculations using 
$E_T$ as renormalization scale. The agreement is good,
although 
NLO pQCD fails to describe the data in the forward region ($\eta_{lab}>1.5$),
when both $E_T$ and $Q^2$ are small. This region corresponds
to the largest NLO/LO corrections.
Figure ~\ref{fig1} shows the inclusive jet cross section $d\sigma/ dE_{T}$
for $1.5<\eta_{lab}<2.8$ as a function
of the transverse energy in the Breit frame $E_{T}$, 
in different regions of $Q^2$. 

Especially at $5<E_T<20$~GeV and $5<Q^2<10~\mbox{GeV}^2$ 
the renormalization scale uncertainties
do not cover the uncertainties of the data.
These theory uncertainties are
estimated by a variation of the scale by a factor of 2 (0.5).

Both the large differences between NLO and LO calculations and the poor
agreement between data and NLO shows the necessity of further theoretical 
progress. 

\begin{figure}
 \epsfig{file=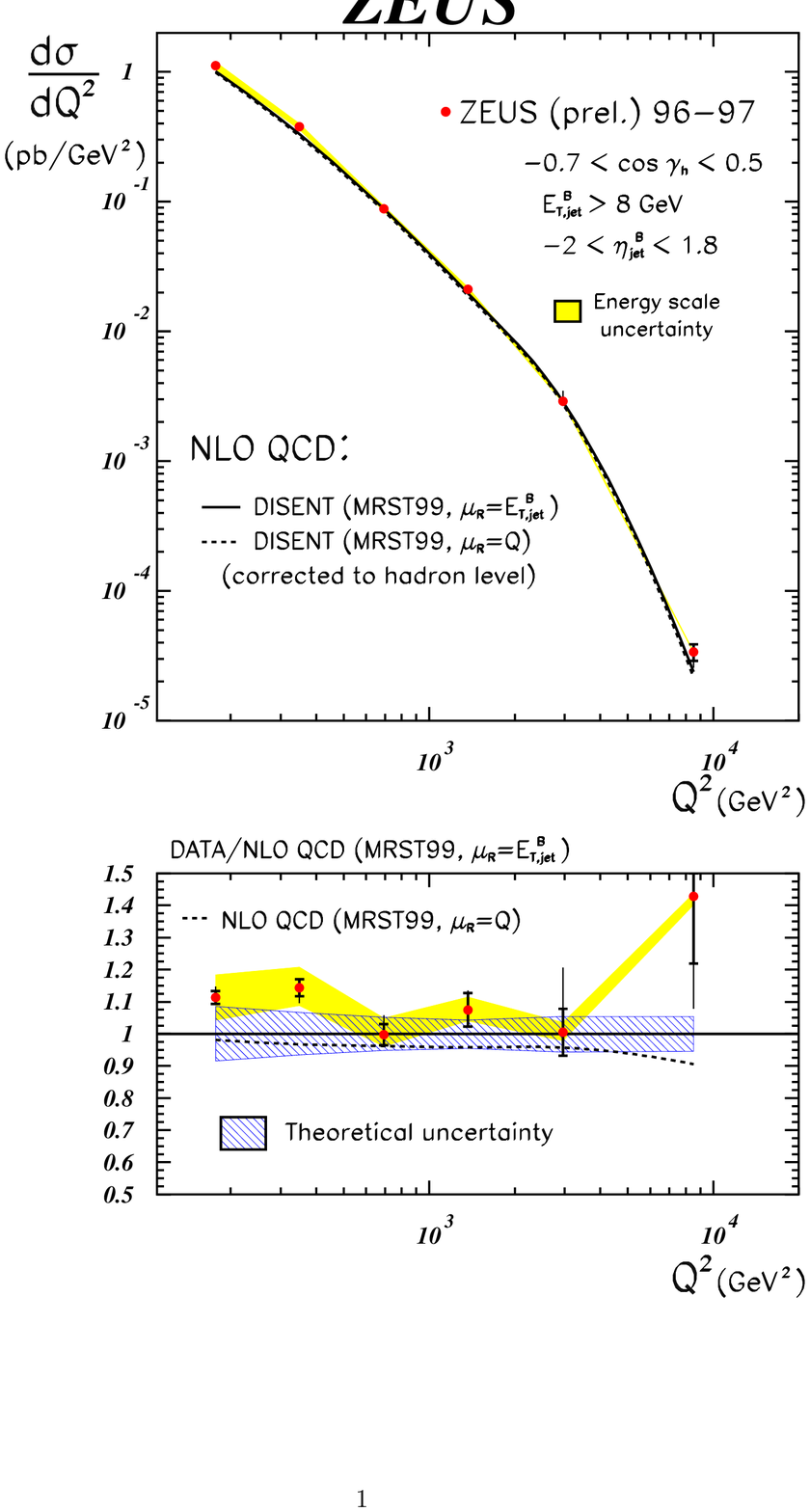,width=6.5cm}
\caption
{Differential
 $ep$ cross section $d\sigma/dQ^2$ for inclusive jet production.
}
\label{fig2}
\end{figure}

As an example of the many jet measurements at high $Q^2$
a new ZEUS measurement of 
inclusive jet cross sections~\cite{Chekanov:2002be} 
for photon virtualities $Q^2>125~\mbox{GeV}^2$
is discussed.
A good agreement between data and NLO calculations
is found at high $Q^2$.
Figure ~\ref{fig2} shows  the differential cross section
$d \sigma / dQ^2$ as a function of $Q^2$ for inclusive jet production
requiring $E_{T}>8$~GeV in the Breit frame.
Since there is good agreement between data and the NLO calculation the 
measured cross sections were used to determine $\alpha_s(M_Z)$.
The QCD fit to H1 and ZEUS high $Q^2$ jet data yield 
$\alpha_s$ values in good agreement with the world average and 
of similar precision to that obtained by other experiments.

\begin{figure}
  \epsfig{file=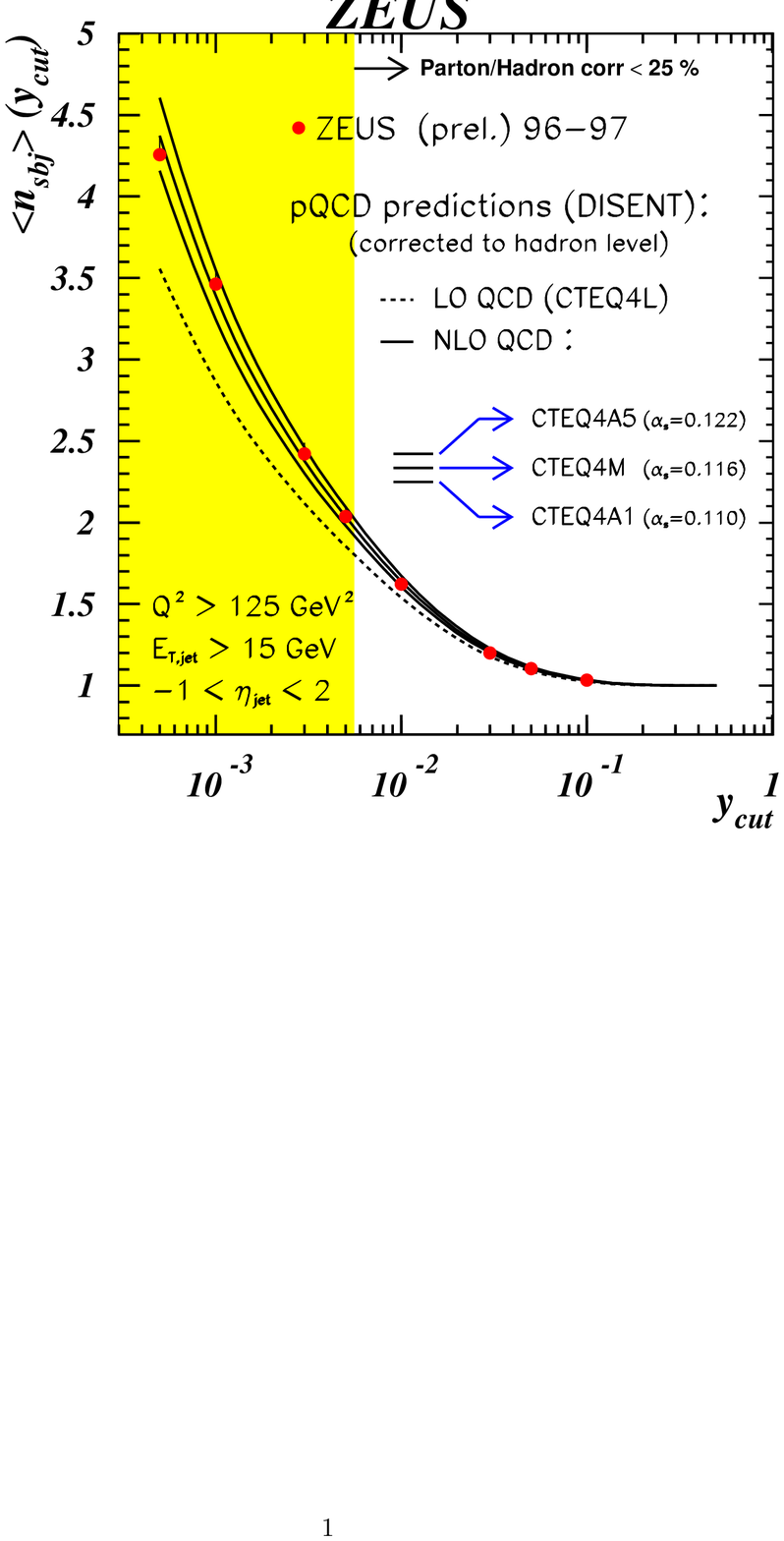,width=7.5cm}
\caption
{The measured mean subjet multiplicity as a function of $y_{cut}$.
}
\label{shape}
\end{figure}


\section{Internal jet structure}
Observables which give insight into the internal jet structure are 
the mean subjet multiplicity and the integrated jet shape.
At high $E_T$, where fragmentation effects are small, these 
observables are calculable in pQCD. 
The subjet multiplicity $<n_{subjet}>$, for instance, is derived by
repeating the $k_T$ jet algorithm
with a smaller resolution scale $y_{cut}$ considering particles lying within 
a jet and counting the number 
of subjets found inside it.
ZEUS has compared new measurements~\cite{zeus1} of these observables to QCD 
calculations at NLO. 
The jets are found in the laboratory frame. 
Figure~\ref{shape} shows the mean subjet multiplicity 
as a function of $y_{cut}$ for $E_T>15~\mbox{GeV}$ and $Q^2>125~\mbox{GeV}^2$.
The NLO pQCD prediction reproduces the data well. This shows, for the first 
time, the ability of NLO pQCD in describing the internal jet structure.
A QCD fit of these data were used to derive $\alpha_s$ values, consistent
with the world average.

\section{Jets in photoproduction}
\begin{figure}
\epsfig{file=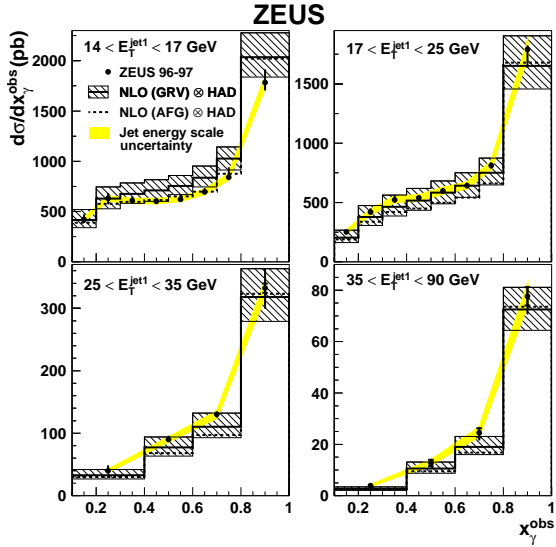,width=7.5cm}
\caption
{Differential $ep$ cross section for dijet photoproduction as a function
of $x_{\gamma}$ for different ranges of the $E_T$ of the leading jet
$E_T^{jet1}$.
}
\label{fig.xgam}
\end{figure}

Dijet photoproduction has been investigated
by the H1 and ZEUS experiments~\cite{Chekanov:2001bw,Adloff:2002au}.
Both analyses present dijet cross sections as functions of
various jet observables.
Asymmetric cuts on the $E_T$ of the two jets are applied in order
to avoid regions of phase space which are 
infrared sensitive in NLO calculations.
Thus the H1 (ZEUS) analysis requires $E_T$ for the first and second jet
of $25$ and $15$ GeV ($14$ and $11$ GeV).
In both analyses 
the ratio of the measured cross sections to the theoretical prediction
varies at small $x_{\gamma}$ by $\pm$10-15\% 
when the cut on $E_{T}$ of the second jet is varied by $\pm 5$~GeV.
This effect is covered by the NLO 
scale uncertainty at low $x_{\gamma}$ of $\approx 15-20$\%.

Figure~\ref{fig.xgam} shows the dijet cross section, as measured 
by ZEUS, as a function of $x_{\gamma}$, which is
an estimator of the momentum fraction carried by partons out of the photon.

Although the shape of the data for different $E_T$ is not perfectly modelled 
by the NLO prediction, strong conclusions can't be drawn, because of the
size of the theoretical uncertainties.
These data are
nevertheless sensitive enough to constrain the photon pdfs (quarks and gluon).
Once more further theoretical understanding would be helpful 
in the interpretation of the data. 

\begin{figure}
 \epsfig{file=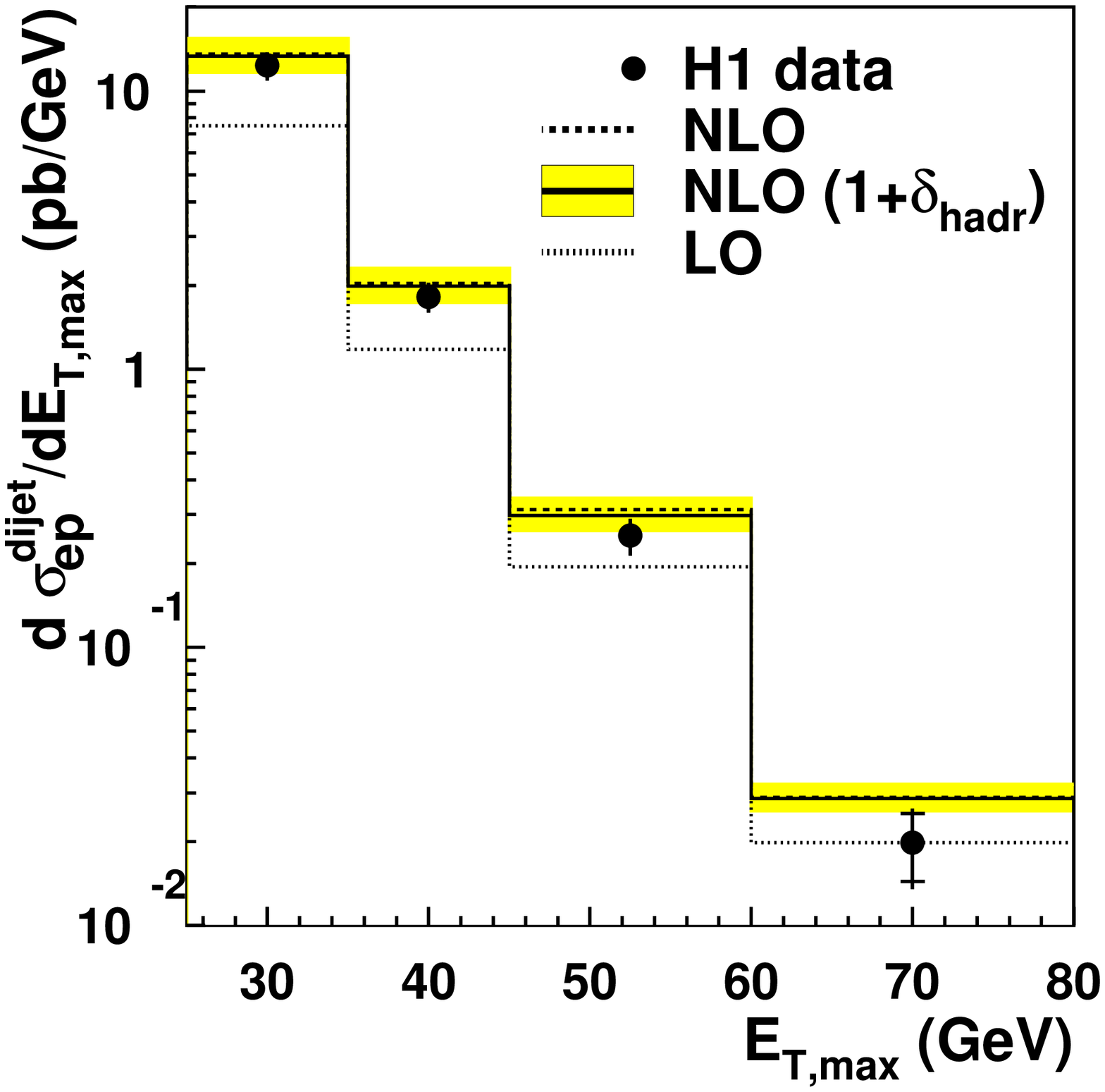,width=6cm}
 \epsfig{file=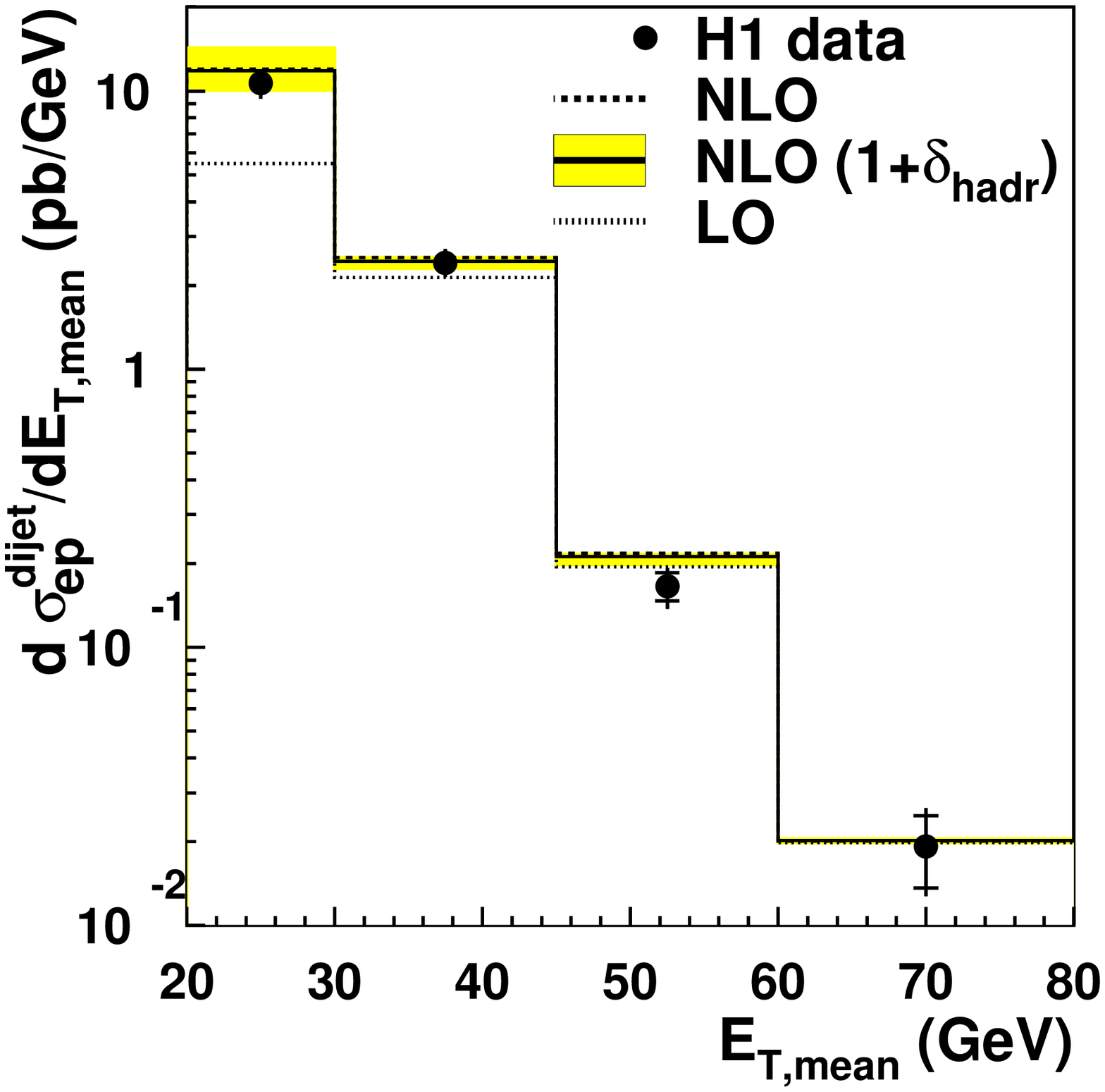,width=6cm}
\caption
{Differential $ep$ cross section for dijet photoproduction as a function
of $E_{T,max}$ (upper Figure) and $E_{T,mean}$ (lower Figure).
}
\label{fig.et}
\end{figure}

In Figure~\ref{fig.et} the H1
dijet cross section as a function of the mean transverse energy of
the two jets 
$d\sigma / d E_{T,mean}$ and of the transverse energy
of the leading jet
$d\sigma / d E_{T,max}$ 
is shown. 
The data are well described up to 
$E_{T}\approx 70 \mbox{GeV}$ by the NLO calculation.  
The NLO scale uncertainty is not reduced significantly
with increasing $E_{T,max}$, but decreases
from $\pm 20$\%
for the first bins to less than $\pm 5$\% for increasing $E_{T,mean}$.
This shows the sensitivity to the
choice of dijet $E_T$ cuts.

\begin{figure}
\epsfig{file=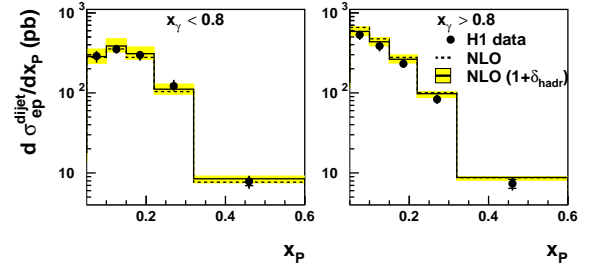,width=8.cm}
\caption
{Differential $ep$ cross section for dijet photoproduction as a function
of $x_P$ for two regions of $x_{\gamma}$.
}
\label{fig.xp}
\end{figure}

The momentum fraction carried by partons out of the proton
is termed $x_P$.
Figure ~\ref{fig.xp} shows 
the cross section $d\sigma / d x_p$
as a function of $x_p$
for two different $x_{\gamma}$ regions.

Even at the highest $x_p$ the measured cross sections
agree well with the QCD predictions.
In this part of the phase space about 40\% of the
cross section is attributed to
processes induced by gluons in the proton.
The recent data could be used to further
constrain the existing proton pdfs, particularly at medium $x_P$.

\section{Summary}
Recent HERA jet data show that 
NLO QCD calculations and current pdf parameterisations
are mostly compatible with high $E_T$ and high $Q^2$ jet data, which 
allows precise determinations of $\alpha_s(M_Z)$ and pdfs.
The increasing precision of the data shows that 
a deeper understanding of jet physics 
will require theoretical progress.


\begin{thebibliography}{9}
\bibitem{jet}
S.~D.~Ellis and D.~E.~Soper,
Phys.\ Rev.\ D {\bf 48} (1993) 3160
[arXiv:hep-ph/9305266].
 
S.~Catani, Y.~L.~Dokshitzer, M.~H.~Seymour and B.~R.~Webber,
Nucl.\ Phys.\ B {\bf 406} (1993) 187.

\bibitem{Adloff:2002ew}
C.~Adloff {\it et al.}  [H1 Collaboration],
`Measurement of inclusive jet cross-sections in deep-inelastic e p  scattering at HERA,''
Phys.\ Lett.\ B {\bf 542} (2002) 193
[arXiv:hep-ex/0206029].


\bibitem{Chekanov:2002be}
S.~Chekanov {\it et al.}  [ZEUS Collaboration],
``Inclusive jet cross sections in the Breit frame in neutral current deep  inelastic scattering at HERA and determination of alpha(s),''
arXiv:hep-ex/0208037.

\bibitem{Chekanov:2001bw}
S.~Chekanov {\it et al.}  [ZEUS Collaboration],
``Dijet photoproduction at HERA and the structure of the photon,''
Eur.\ Phys.\ J.\ C {\bf 23} (2002) 615
[arXiv:hep-ex/0112029].

\bibitem{Adloff:2002au}
C.~Adloff {\it et al.}  [H1 Collaboration],
``Measurement of dijet cross sections in photoproduction at HERA,''
Eur.\ Phys.\ J.\ C {\bf 25} (2002) 13
[arXiv:hep-ex/0201006].

\bibitem{zeus1}
S.~Chekanov {\it et al.}  [ZEUS Collaboration],
``Measurements of jet substructure in
neutral current deep inelastic scattering
and determination of $\alpha_s$ at HERA'' 
, EPS01 conference, Abs 641

\end{thebibliography}
\end{document}